\documentclass[10pt,letterpaper]{article}
\usepackage{opex3}
\usepackage[]{graphicx}
\usepackage[]{amsmath}
\usepackage{epstopdf}

\usepackage{ae} %%for Computer Modern fonts

\begin{document}

%%%%%%%%%%%%%%%%%% title page information %%%%%%%%%%%%%%%%%%
\title{Dispersion properties of a~nanophotonic Bragg waveguide with finite aperiodic cladding}

\author{Volodymyr~I.~Fesenko$^{1,*}$, Vladimir~R.~Tuz$^{1,2}$, Oleksiy~V.~Shulika$^3$, and Igor~A.~Sukhoivanov$^3$}

\address{$^1$Institute of Radio Astronomy of NASU, Ukraine \\
$^2$School of Radio Physics, V.N. Karazin Kharkiv National University, Ukraine \\
$^3$Department of Electronic Engineering and Communications, DICIS,
University of Guanajuato, Mexico}

\email{$^*$v.i.fesenko@ieee.org; volodymyr.i.fesenko@gmail.com} %% email address is required


\begin{abstract}  A comprehensive
analysis of guided modes of a novel type of a planar Bragg
reflection waveguide which consists of a low refractive index
guiding layer sandwiched between two finite aperiodic mirrors is
presented. The layers in the mirrors are aperiodically arranged
according to the Kolakoski substitution rule. In such a waveguide
light is confined inside the core by Bragg reflection from the
mirrors, while dispersion characteristics of guided modes strongly
depend on aperiodicity of the cladding. Using the transfer matrix
formalism bandgap conditions, dispersion characteristics and mode
profiles of the guided modes of such Bragg reflection waveguide are
studied.
\end{abstract}

\ocis{(130.3120) Integrated optics devices; (130.5296) Photonic
crystal waveguides; (230.7390) Waveguides, planar; (230.4170)
Multilayers; (310.4165) {Multilayer design}.}

%%%%%%%%%%%%%%%%%%%%%%% References %%%%%%%%%%%%%%%%%%%%%%%%%

%%%%%%%%%%%%%%%%%%%%%%%%%%  body  %%%%%%%%%%%%%%%%%%%%%%%%%%
\section{Introduction}
\label{sec:intro}

Bragg reflection waveguides \cite{fox_ProcIEEE_1974,
yeh_OptCommun_1976} are photonic structures, designed to guide light
within a core layer surrounded by a special composite cladding whose
effective refractive index can be either lower or higher then the
core one. The cladding can have different configurations, e.g. it
can be designed as a multilayer stack of periodically arranged
dielectric layers (slab waveguides) \cite{fox_ProcIEEE_1974,
yeh_OptCommun_1976}, coaxial layers having alternated high and low
refractive indices (Bragg fibers) \cite{yeh_JOSA_1978, hu_JNN_2013},
arrays of holes in a single-material dielectric film or fiber
\cite{sukhoivanov_OE_2013}, systems of dielectric pillars
\cite{Chigrin_OE_2004}, etc. In all mentioned designs, unlike the
conventional high-index guiding waveguides based on the total
internal reflection \cite{snyder_1983}, principal characteristics of
Bragg reflection waveguides are influenced by periodicity in their
cladding constituents. Indeed, the periodicity in constituents leads
to formation of photonic bandgaps in the spectra of Bragg mirrors
resulting in light confinement within a core layer. Such photonic
bandgap guidance brings several attractive features to the waveguide
characteristics \cite{russel_Science_2003}, in particular, since
most of light is guided inside a low-index core (which can even be
an air channel), losses and nonlinear effects can be significantly
suppressed. Moreover, the mode area, mode profile and dispersion
properties of Bragg reflection waveguides can be optimized providing
specific choice of constituents and cladding configuration(e.g.
utilization of chirping \cite{nistad_OptCommun_2006} or aperiodic
\cite{fesenko_SPIE_2014} designs for the Bragg mirrors). The
interest to the mentioned unique features of Bragg reflection
waveguides arises sufficiently in recent years due to advances in
the deposition and crystal growth technologies, which made possible
the fabrication of waveguides with complicate designs assuring an
appropriate quality.

First of all, as an ideal model, a Bragg reflection waveguide made
of a low-index layer sandwiched between two identical {\it
semi-infinite} one-dimensional Bragg mirrors is considered in
\cite{yeh_OptCommun_1976}. In such a system the semi-infinite
configuration of mirrors guarantees existence of the total
reflection within complete photonic bandgaps that provides perfect
light guiding inside the core. From the theoretical standpoint, an
efficient mathematical method based on the Bloch wave formulation is
applied to analyze and tailor the propagation characteristics of
modes in such waveguide. However, in designs of practical systems
the multilayer structure of cladding definitely has a {\it finite}
extent which requires taking into account an energy leakage from the
waveguide through its imperfect mirrors that is out of scope of the
Bloch wave formulation related to the infinitely extended periodic
medium.

In order to consider a more realistic model of Bragg reflection
waveguides with a finite extent of multilayer mirrors among other
purely numerical simulations based on propagation methods
\cite{li_OptCommun_2008, fesenko_intech_2013}, in the
one-dimensional case the transfer matrix formalism
\cite{dasgupta_OptCommun_2007} is widely used which assumes matching
the boundary conditions at interfaces between high- and low-index
layers and enables one to derive an analytic solution on this basis.
Although the latter method is only applicable for simple
configurations of Bragg reflection waveguides, in many cases it
allows to offer deep physical insight into the problem. The solution
for a planar Bragg waveguide can be derived in the form of the TE
and TM modes considered separately, with two sets of $2 \times 2$
transfer matrices, while the cylindrical Bragg fiber case requires
$4 \times 4$ matrices in general. Nevertheless, since most of
characteristics of planar Bragg waveguides are similar to those of
cylindrical Bragg fibers, it is well known that results obtained
with the transfer matrix method for one-dimensional planar systems
are also applicable for prediction of optical features of fibers
without re-solving the problem for the cylindrical geometry.

In the search for improved features of Bragg reflection waveguides,
aside from a standard configuration of the waveguide with a periodic
quarter-wave cladding \cite{west_JOSAB_2006}, it has been proposed
and studied a number of designs among which we distinguish
particular configurations consisted of a guiding layer placed
between two different Bragg reflectors \cite{li_OptCommun_2008,
li_OptExpress_2009}, mirrors with a periodicity defect
\cite{nistad_OptCommun_2006} and chirped claddings
\cite{nistad_OptCommun_2006, Pal_OptQuantElectron_2007}. Thus it is
shown that in the case of a waveguide with two different Bragg
reflectors two distinct sets of bandgaps can be obtained and the
waves guided along the waveguide appear to be inside their
overlapped bandgaps. By manipulating two sets of bandgaps
independently, the flexibility in the control of transmission
characteristics of the guided modes can be enhanced significantly.
Furthermore by introducing a defect inside the mirror's structure it
is possible to shift the zero-dispersion point toward longer
wavelengths.

At the same time the most promising technique is seen to be in the
use of the omnidirectional reflection feature which is available in
multilayer photonic structures \cite{Fink_Science_1998}. In
\cite{ibanescu_Science_2000, johnson_ProcSPIE_2002} a particular
design of Bragg reflection waveguides utilizing omnidirectional
reflection is referred to ``OmniGuide" fibers. Such omnidirectional
reflection in the waveguide cladding results in strong resemblances
of OmniGuide fiber features to those ones of the convenient hollow
metallic waveguides, confining a set of guided modes almost entirely
within the hollow core with similar field patterns and dispersion
characteristics.

Here we propose a way for the integration of these mentioned
improvements within a single system by utilization of a
deterministically {\it aperiodic}  design in the cladding. It is no
doubt that a deterministically aperiodic multilayer system can be
considered as a periodic one having plural defects and thus
aperiodic configuration gives an opportunity  to customize
dispersion characteristics of the waveguide. Another key point is
that aperiodic multilayers possess a much more complex structure in
the reciprocal space than periodic ones. It results in presence of
several omnidirectional photonic bandgaps within a period of the
reciprocal space of aperiodic systems \cite{lusk_OptCommun_2001,
qiu_EPL_2004, barriuso_OptExpress_2005}.

From our previous studies \cite{fesenko_SPIE_2013,
fesenko_WRCM_2014, fesenko_PIERM_2015} it is found that
deterministically aperiodic multilayers constructed according to the
Kolakoski generation rules demonstrate pronounced omnidirectional
reflectance and have completely transparent states in the
transmission spectra which are important for achieving sufficient
waveguide modes filtering. Therefore, in this study, we introduce an
extra degree of freedom for optimizing Bragg reflection waveguides
through utilization of an aperiodically layered cladding. Our goal
here is to study the dispersion characteristics providing the mode
classification of a planar waveguide system whose mirrors are made
from dielectric layers arranged according to the mentioned Kolakoski
generation rules.  The solution for both TE and TM modes is
constructed using the transfer matrix formalism \cite{Tuz_JOSA_09,
Tuz_WRCM_09, Tuz_JOpt_09}.

%%%%%%%%%%%%%%%%%%%%%%%%%%%%%%%%%%%%%%%%%%%%%%%%%%%%%%%%%%%%%
\section{Theoretical Description}
\label{sec:problem}

\subsection{Aperiodic Bragg Reflection Waveguide Configuration}
\label{sec:configuration}

\begin{figure}[h]
\begin{center}
\begin{tabular}{c}
\includegraphics[height=6.0cm]{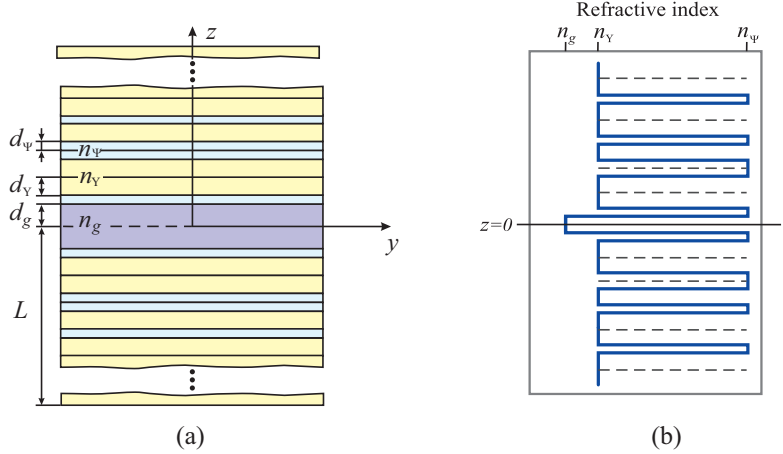}
\end{tabular}
\end{center}
\caption[example] {\label{fig:fig_1} (a) The schematic of a Bragg
reflection waveguide with aperiodic mirrors arranged according to
the Kolakoski $K(1, 2)$ substitutional rules; (b) index profile of
the structure.}
\end{figure}

Further we consider a Bragg reflection waveguide made of low-index
core layer sandwiched between two identical aperiodic
one-dimensional mirrors formed by stacking together two different
materials $\Psi$ and $\Upsilon$ according to the Kolakoski $K(p, q)$
generation scheme as it is shown in Fig.~\ref{fig:fig_1}a.

The generation rule $w$ of the Kolakoski sequence is similar to
those ones of the Fibonacci or Thue-Morse sequences and can be based
on two symbols substitution. Namely the sequence $K(p, q)$ can be
obtained by starting with $p$ as a seed and iterating the following
two substitutions:
\begin{equation}
    \label{eq:Kpq}
    w_0~:
    \begin{array}{ccccccccccccc}
          q~\to p^{q}\\
          p~\to p^{p}\\
    \end{array}
~~~~~~~\mbox{and}~~~~~~~~w_1:
    \begin{array}{ccccccccccccc}
          q~\to q^{q}\\
          p~\to q^{p}\\
    \end{array}
\end{equation}
where $w_0$ and $w_1$ can be any string of letters $p$ and $q$;
$p^q$ denotes a run of $p$ $q$'s, i.e., $p^q = p...p$ ($q$ times).
Extra information about generation rules of the Kolakoski sequence,
its general properties and configuration of different multilayers
based on the Kolakoski generation scheme can be found in Refs.
\cite{Kolakoski65, Sing04} and Refs. \cite{fesenko_SPIE_2013,
fesenko_WRCM_2014, fesenko_PIERM_2015}, respectively. In this paper
the number of generation stage of the sequence is defined as
$\sigma$.

The structure under study is infinite along $x$-axis, so
$\partial/\partial x=0$. Besides, the refractive index profile of
the waveguide structure varies only along $z$-axis and it is
invariant in other directions (see, Fig.~\ref{fig:fig_1}b). In this
configuration the middle of the core is assumed to be at the line $z
= 0$. We suppose that both mirrors consist of a finite number $N$ of
the constitutive layers $(\sigma < \infty)$. Also it is further
assumed that the letters $\Psi$ and $\Upsilon$ denote two different
layers with thicknesses $d_\Psi$, $d_\Upsilon$ and refractive
indices $n_\Psi$, $n_\Upsilon$, respectively. The guiding layer has
thickness $2d_g$ and refractive index $n_g$. The total thickness of
the structure under study is $2L$. In general case, the external
medium outside the layered system $z<-L$ and $z>L$ is homogeneous,
isotropic and can have either high or low refractive index, however,
in this paper we assume that the waveguide is surrounded by air with
$n_0=1$.

In the chosen structure configuration, each guided mode propagates
along $y$-axis with its own propagation constant $\beta$. In our
study we analyze the propagation of both TE waves and TM waves with
field components $\vec E = \{E_x, 0, 0\},~\vec H=\{0, H_y, H_z\}$
and $\vec E = \{0,E_y,E_z\},~\vec H=\{H_x, 0, 0\}$, respectively.
Note that in the case of the practical Bragg reflection waveguides
with finite number of layers in the mirrors, the light propagation
is accompanied by some attenuation that appears as a result of the
energy leakage through imperfect mirrors, which is consistently
taken into consideration within our model.

%%-----------------------------------------------------------
\subsection{Transfer Matrix Description}
\label{sec:TMM}

Key task here is a definition of dispersion fields of the finite
aperiodic sequence of alternating two different isotropic layers.
The plane monochromatic waves of TE ($\vec E^{\mathrm{TE}} \parallel
\vec x_0 $) and TM ($\vec H^{\mathrm{TM}} \parallel \vec x_0 $)
polarizations can be defined in a particular $j$-th layer
($j=1,2,...,N)$ of the sequence in the form
\begin{equation}
\label{eq:fieldcomponents} \left\{ \begin{matrix} \vec
E^{\mathrm{TE}}_j \\ \vec H^{\mathrm{TM}}_j
\end{matrix}\right\}=\vec x_0\left\{ \begin{matrix}
1/\sqrt{Y_j^{\mathrm{TE}}} \\ \sqrt{Y_j^{\mathrm{TM}}}
\end{matrix}\right\}u_j(z)\exp[-i(\omega t - \beta y)],
\end{equation}
where
\begin{equation}
\label{eq:amplitude} u_j(z)=a_j\exp(ik_{zj}z)+b_j\exp(-ik_{zj}z),
\end{equation}
and $a_j$ and $b_j$ are the field amplitudes,
$Y_j^{\mathrm{TE}}=k_{zj}/k_0\mu_j$ and
$Y_j^{\mathrm{TM}}=k_0\varepsilon_j/k_{zj}$ are the wave
admittances,  $k_0=\omega/c$ is the wavenumber in free space, and
$k_{zj}$ is the transverse wavenumber which takes on discrete values
in each slab and can be written as
\begin{equation}
\label{eq:wavenumber} k_{zj}=\left(k_0^2n_j^2
-\beta^2\right)^{1/2}=k_0\left(n_j^2-n_{eff}^2\right)^{1/2}.
\end{equation}
Here $n_{eff}=\beta/k_0$ is introduced as the effective refractive
index of the guided mode, $n_j$ takes on values $n_g$ for the core
layer, and $n_\Psi$ and $n_\Upsilon$ for the $\Psi$ and $\Upsilon$
cladding layers, respectively.

The field amplitudes for the structure input and output are
evaluated as \cite{Tuz_JOSA_09, Tuz_WRCM_09, Tuz_JOpt_09}
\begin{equation}
\label{eq:matr1}
\begin{array}{ccccccccccccc}
\left[ \begin{matrix}a_{0}\\b_{0}\end{matrix}\right] =
\mathbf{T}_\Sigma\left[
\begin{matrix}a_{N}\\b_{N}\end{matrix}\right] = &
(~\underbrace{\mathbf{T}_\Psi \mathbf{T}_\Upsilon
\mathbf{T}_\Upsilon \mathbf{T}_\Psi \mathbf{T}_\Psi
\mathbf{T}_\Upsilon\ldots }~)& \left[
\begin{matrix}a_{N}\\b_{N}\end{matrix}\right],
\\
           & N &
\end{array}
      \end{equation}
where the total transfer matrix $\mathbf{T}_\Sigma$ is obtained by
multiplying in the appropriate order the matrices corresponding to
each layer in the structure.

The matrices $\mathbf{T}_\Psi$ and $\mathbf{T}_\Upsilon$ are the
particular transfer matrices of rank 2 of the $\Psi$ and $\Upsilon$
layers in cladding with their corresponding thicknesses $d_\Psi$ and
$d_\Upsilon$. They are
\begin{equation}
\label{eq:transferUP}
\mathbf{T}_\Psi=\mathbf{T}_{01}\mathbf{P}_1(d_\Psi)\mathbf{T}_{10},~~~
\mathbf{T}_\Upsilon=\mathbf{T}_{02}\mathbf{P}_2(d_\Upsilon)\mathbf{T}_{20},
\end{equation}
where $\mathbf{T}_{0j}$ and $\mathbf{T}_{j0}$ ($j = 1, 2$) are the
transfer matrices of the layer interfaces with outer half-spaces,
and $\mathbf{P}_j(d)$ are the propagation matrices through the
corresponding layer. The elements of the matrices $\mathbf{T}_{0j}$
and $\mathbf{T}_{j0}$ are determined by solving the boundary-value
problem related to the field components (\ref{eq:fieldcomponents}):
\begin{equation}
\label{eq:matr01} \mathbf{T}_{pv} = \frac{1}{2\sqrt{Y_pY_v}}\left[
\begin{matrix}
Y_p+Y_v & \pm( Y_p-Y_v) \\
\pm( Y_p-Y_v) & Y_p+Y_v
\end{matrix}
\right],~~~ \mathbf{P}_{j}(d) = \left[
\begin{matrix}
\exp(-ik_{zj}d) & 0 \\
0 & \exp(ik_{zj}d)
\end{matrix}
\right],
\end{equation}
where the upper sign `$+$' relates to the TE wave, while the lower
sign `$-$' relates to the TM wave.

Finally, the reflection coefficient of the layer stack is determined
by the expression
\begin{equation}
\label{eq:Ref}
 R= (b_0/a_0)|_{b_N=0}=-t_{21}/t_{22},
 \end{equation}
where $t_{mn}$ are the elements of the total transfer matrix
$\mathbf{T}_\Sigma$.

\subsection{Dispersion Equation for Guided Modes }
\label{sec:DispRel} As Bragg mirrors on both sides of the waveguide
core layer are the same (i.e. the Bragg reflection waveguide is
symmetrical about the $z$-axis, $n(-z) = n(z)$), the equations for
waves travelling back and forth inside the channel can be joined on
the boundaries $z = d_g$ and $z=-d_g$ into the next system
\begin{equation}
      \label{eq:dispersionsyst}
\left\{
\begin{aligned}
&a_0\exp[-ik_{zg}d_g]=Rb_0\exp[ik_{zg}d_g],\\
&b_0\exp[-ik_{zg}d_g]=Ra_0\exp[ik_{zg}d_g],
\end{aligned}
\right.
\end{equation}
from which the relation between amplitudes can be found
\begin{equation}
      \label{eq:dispersionamp}
b_0=a_0R\exp[2ik_{zg}d_g].
\end{equation}
Eliminating amplitudes $a_0$ and $b_0$ from system
(\ref{eq:dispersionsyst}) the {\it dispersion} equation for the
guided modes of the Bragg reflection waveguide is obtained as
\begin{equation}
      \label{eq:dispersion}
1-R^2\exp\left [4ik_0d_g\sqrt{n_g^2-n_{eff}^2}~\right]=0.
\end{equation}
This equation is further solved numerically to find out a function
of the propagation constant $\beta$ versus frequency $\omega$.
Remarkably, even in the case when constitutive materials of layers
in the core and cladding are considered to be without intrinsic
losses, for a finite number of these layers $N$ in the multilayer
system, solutions of the dispersion equation (\ref{eq:dispersion})
appear in the field of complex numbers $\beta$ since there is an
inevitable energy leakage through the outermost layers. Thus, the
resulting propagation constant is sought in the complex form as
$\beta=\beta'+i\beta''$.

In a semi-infinite periodic structure without intrinsic losses in
its constituents, the propagation constant $\beta$ takes either
purely real ($\beta''=0$) or purely imaginary ($\beta'=0$) values
\cite{yeh_OptCommun_1976}, where the real $\beta$ corresponds to
propagating waves. In the case of structure with a finite number of
layers in the cladding, the roots of the dispersion equation are
always complex numbers, nevertheless in order to provide the wave
propagation, the imaginary part of $\beta$ must be small enough. As
it was demonstrated in \cite{west_JOSAB_2006}, $\beta''$ becomes to
be negligibly small for the number of cladding's layers more then $N
= 24$ for the chosen structure parameters confirming that losses
reach a value lower than $10^{-4}$~dB/cm in their case. Thus, in our
consideration the number of layers in the cladding is defined in
such a way to satisfy the criterion of smallness of the imaginary
part of $\beta$ when classifying waveguide modes. All other
solutions are interpreted as those that belong to leaky waves and
therefore are discarded from our consideration.

Furthermore, for the symmetric waveguides all guided modes can be
divided onto symmetric and anti-symmetric ones. Without loss of
generality, the overall field amplitude in the middle line of the
core can be reduced to unity or zero for symmetric and
anti-symmetric modes, respectively. Then the field in the core can
be normalized by setting $a_0=1/2$ and $b_0=\pm1/2$, where the upper
sign `$+$' is related to symmetric modes, while the lower sign `$-$'
is related to anti-symmetric ones, respectively.

%%%%%%%%%%%%%%%%%%%%%%%%%%%%%%%%%%%%%%%%%%%%%%%%%%%%%%%%%%%%%
\section{Numerical Results: Solution Analysis}
\label{sec:results}

\subsection{Spectral Properties and Dispersion Diagrams}
\label{sec:dispersion}

For a comparative study we further consider two different
configurations of a Bragg reflection waveguide. The first
configuration is a waveguide with a typical periodic cladding. In
this case we repeat the results of \cite{I_JKPS2001} in order to
verify the method of solution and comply the succession. In the
second configuration the cladding is considered to be an aperiodic
structure arranged according to $K(1,2)$ generation scheme. Thus, in
both configurations we study a waveguide having an air core layer
with thickness $2d_g=(2/3)(d_\Psi + d_\Upsilon)$ and refractive
index $n_g =1.0$. The air guiding layer is sandwiched between two
identical one-dimensional photonic structures which generally
reflect light back over a bounded range of angles. Nevertheless, in
the particular case of existing of the omnidirectional reflection,
the angular range is complete \cite{Fink_Science_1998}. As
constituents for cladding we consider combination of layers made of
GaAs and oxidized AlAs, whose refractive indices at the wavelength
of $1~\mu\mathrm m$ are $n_\Psi = 3.50$ and $n_\Upsilon = 1.56$,
respectively. The thickness ratio is defined to be $d_\Psi :
d_\Upsilon = 1 : 2$. Note that similar results can be obtained by
choosing alternation of layers made of Si and SiO${_2}$, whose
refractive indices at the wavelength of $1~\mu\mathrm m$ are $n_\Psi
= 3.56$ and $n_\Upsilon = 1.46$, respectively. This pair of the
cladding's materials will be further used for the chromatic
dispersion calculation.

\begin{figure}[h]
\begin{center}
\begin{tabular}{c}
\includegraphics[height=5.5cm]{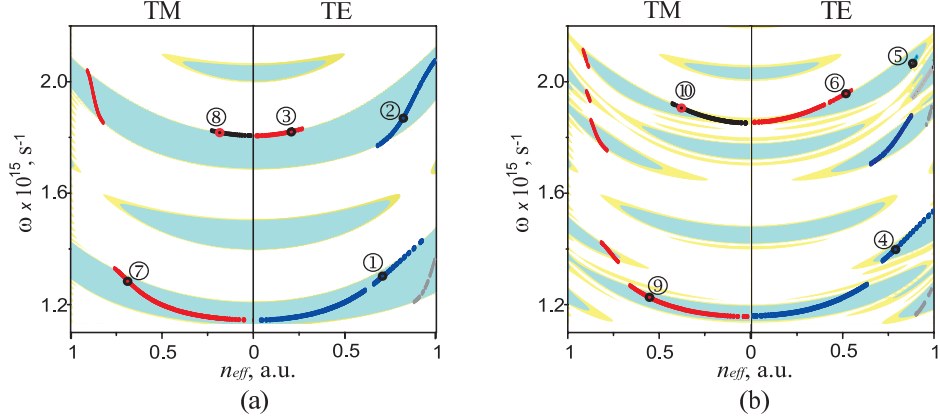}
\end{tabular}
\end{center}
\caption[example] {\label{fig:fig_2} The band diagrams and
dispersion curves for TM and TE waves in the Bragg reflection
waveguide with (a) periodic cladding and (b) aperiodic cladding. The
colored areas correspond to bandgaps with level of reflection $|R|>
0.9$. The different colors of the dispersion curves correspond to
different values of the mode index $m$: blue curves correspond to
$m=0$, red curves correspond to $m=1$, and black curves correspond
to $m=2$. The gray lines represent modes with $m=-1$; $N=24$.}
\end{figure}

As a typical example in Fig.~\ref{fig:fig_2} we demonstrate the band
diagrams and dispersion curves which are calculated for two
mentioned particular waveguide's cladding configurations. These
results are obtained for the same number $N$ and material parameters
$n_\Psi$ and $n_\Upsilon$ of the constitutive layers for both
cladding configurations. Here the regions  colored in light yellow
indicate the photonic bands of the Bragg mirrors where the level of
reflection lies within the range $0.9 < |R| < 0.999$ for both TE and
TM waves. Also the regions colored in light blue correspond to the
level of reflection $|R| \ge 0.999$ which can be identified as
bandgaps. The middle line $n_{eff} = 0$ divides band diagrams onto
two parts, where the left side represents features of TM waves,
while the right one represents features of TE waves.

In the case of periodic cladding (Fig.~\ref{fig:fig_2}a), in the
band diagram on either side of the line $n_{eff} = 0$ there is a set
of $n$ disjoint bands where the level of reflection riches very high
values ($R \ge 0.999$). They are centered around frequencies
$\omega_\nu = \nu\pi c/(n_\Psi d_\Psi + n_\Upsilon d_\Upsilon)$,
where $\omega_\nu$ is a central frequency of the corresponding band
and $\nu$ is the band order (in the presented fragment of the band
diagram $\nu$ acquires values 4, 5, 6 and 7). At the same time the
band diagram of the system with an aperiodic arrangement is more
complicated (Fig.~\ref{fig:fig_2}b). Thus, on both sides of the line
$n_{eff} = 0$ there is a set of bandgaps ($|R| \ge 0.999$) with the
same central frequency $\omega_\nu$ as for the periodic structure.
But all the bands get narrower for both TE and TM waves. Besides, as
value of $n_{eff}$ moves towards 1 some additional high-reflectance
bands appear for both TE and TM waves.

Furthermore in both structure configurations the coverage areas of
bands related to TM waves are smaller than those ones of TE waves,
which is consistent with other observations on the reflection
bandwidths of the two polarizations (see, for instance
\cite{qiu_EPL_2004, fesenko_PIERM_2015}). In fact, the bands of TM
waves appear to be totally covered by the corresponding bands of TE
waves. Therefore, it is enough to provide analysis of
omnidirectional reflection only on the basis of bands
characteristics of TM waves.

In particular, in our previous investigations
\cite{fesenko_SPIE_2014, fesenko_PIERM_2015} it is demonstrated that
multilayer structures, arranged according to the classical and
generalized Kolakoski self-generation schemes, can acquire several
overall omnidirectional reflection bands. The omnidirectional
reflection is achieved by accurate adjusting material parameters and
geometrical structure of the distributed Bragg reflectors, such as
refractive indices ratio $r=n_\Upsilon/n_\Psi$ and thicknesses of
the constitutive layers $d_\Psi$ and $d_\Upsilon$. Particularly, the
refractive indices ratio $r$ should be as high as possible, but for
practical structures this ratio can be limited by  availability of
materials that are used in the integrated optic circuits. Therefore
the thicknesses of the layers also should be accurately tuned,
namely, the individual layer thickness of a surrounded Bragg
reflector must be chosen exactly to be one quarter of an operating
wavelength \cite{I_jlwt2004}. In our case it is obvious that for the
structure under study, these criteria are not met and, therefore,
there is not any omnidirectional reflection band for both structure
configurations.

On the other hand Fig.~\ref{fig:fig_2} presents information about
calculated dispersion curves for both configurations of a Bragg
reflection waveguide having periodic or aperiodic cladding. The
waveguide modes are indicated by a set of colored points. All these
modes appear only within the gaps where the level of reflection
riches high values ($|R| \ge 0.999$). Within the neighbor  bandgaps
dispersion curves can have the same mode index $m$, while they
disappear outside the range of bandgaps. In fact, these curves have
discontinuities at the bandgap edges at which there is a mode
cutoff. Moreover, in the aperiodic configuration the dispersion
curves have more cutoff points for each mode than those of periodic
one. It means that Bragg reflection waveguides with aperiodic
cladding are easier to be optimized than that ones with periodic
cladding to support propagation of desired modes only. Thus, an
aperiodic configuration of cladding of Bragg reflection waveguide
can give rise to exceptionally strong mode selection and tuning the
polarization-discrimination effects.

\subsection{Mode-Field Patterns and Confinement Factor}
\label{sec:fields}

\begin{figure}[h]
\begin{center}
\begin{tabular}{c}
\includegraphics[height=7.0cm]{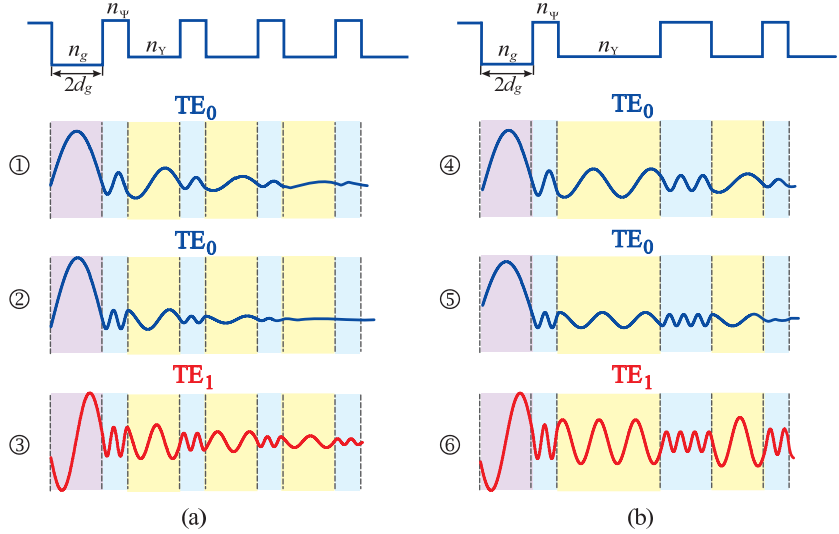}
\end{tabular}
\end{center}
\caption[example] {\label{fig:fig_3} Electric-field patterns related
to the $E_x$ component of the field for TE$_m$ modes. The refractive
index profile of the corresponding waveguide is presented on the top
of figure. The confinement factor is: (1)~$\Gamma=0.77$;
(2)~$\Gamma=0.92$; (3)~$\Gamma=0.22$; (4)~$\Gamma=0.73$;
(5)~$\Gamma=0.87$; (6)~$\Gamma=0.35$.}
\end{figure}

\begin{figure}[h]
\begin{center}
\begin{tabular}{c}
\includegraphics[height=5.5cm]{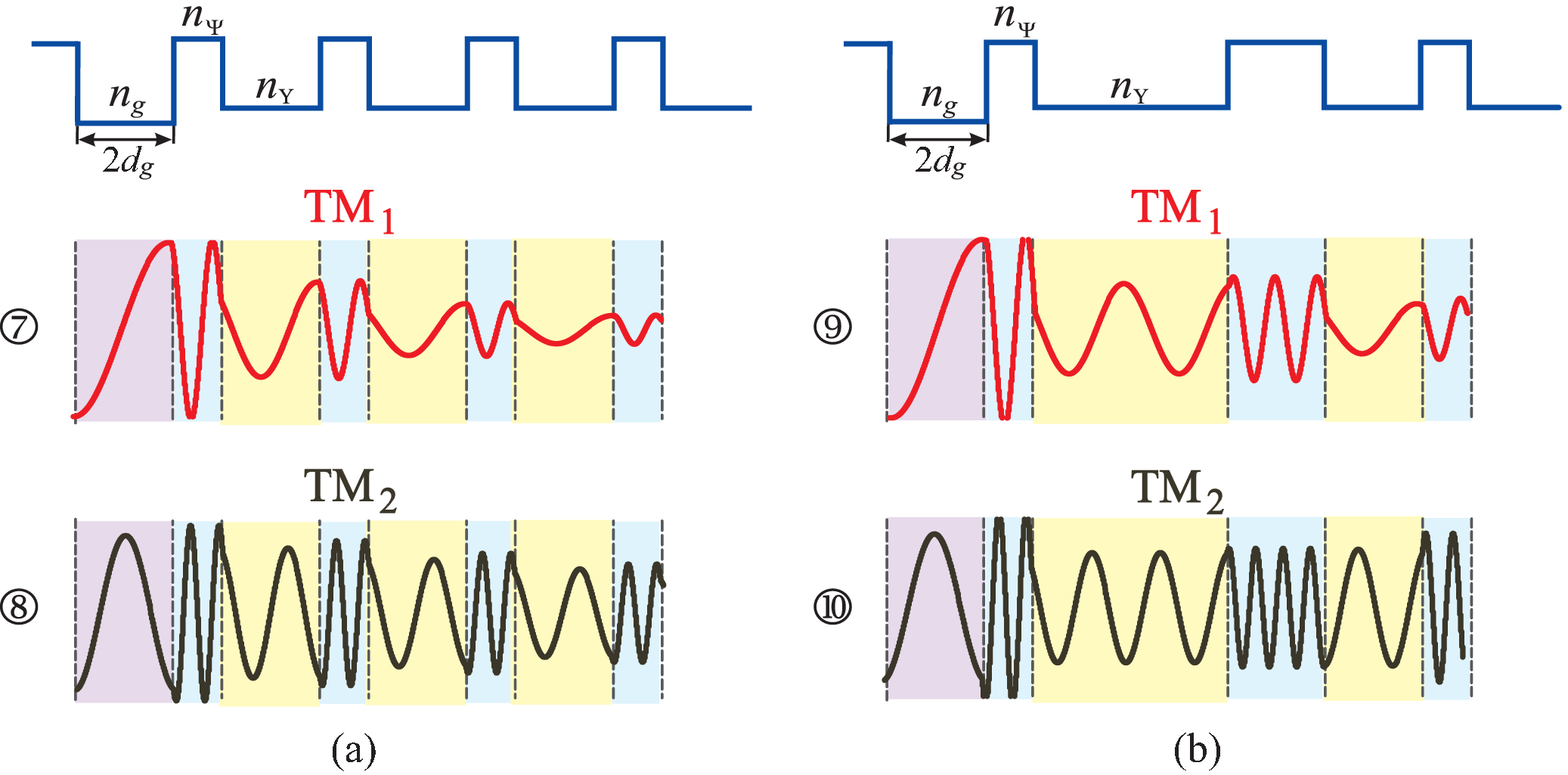}
\end{tabular}
\end{center}
\caption[example] {\label{fig:fig_4} Magnetic-field patterns related
to the $H_x$ component of the field for TM$_m$ modes. The refractive
index profile of the corresponding waveguide is presented on the top
of figure. The confinement factor is: (7)~$\Gamma=0.3$;
(8)~$\Gamma=0.14$; (9)~$\Gamma=0.23$; (10)~$\Gamma<0.1$.}
\end{figure}

A guided mode in the Bragg reflection waveguide can be considered as
a plane wave travelling back and forth in the core, forming a
standing wave pattern. Figs.~\ref{fig:fig_3} and \ref{fig:fig_4}
depict the mode-field patterns for a number of TE$_m$ and TM$_m$
modes with the lowest index $m$. As before in these figures both
periodic and aperiodic configurations of cladding are presented. The
numbers in circles from 1 to 10 are ranged according to the marked
points presented on the dispersion curves in Fig.~\ref{fig:fig_2}.
The patterns are plotted after applying normalization of the field
magnitude on its maximal value within the core. The fractional power
in the guiding layer is known as the energy confinement factor
$\Gamma$ that is defined in the form \cite{I_jlwt2004, Agrawal2001}
\begin{equation}
\label{eq:confinement}
\Gamma=\frac{\int_{-d_g}^{d_g}|\eta|^2\,dz}{\int_{-L}^L
|\eta|^2\,dz},
\end{equation}
where $\eta$ is related to field components $E_x$ and $H_x$ for TE
and TM waves, respectively.
As known in a conventional slab waveguide, with $n_g > n_{clad}$,
the field decays exponentially in the cladding as waves propagate
away from the core. However, in a planar Bragg reflection waveguide
the field acquires some oscillations inside the layered system of
mirrors for both the periodic and aperiodic configurations. The
magnitude of these oscillations decays monotonically away from the
guiding layer as it is depicted in Figs.~\ref{fig:fig_3} and
\ref{fig:fig_4}. Besides, as it is described in
\cite{li_OptCommun_2008, li_josa2007}, for Bragg reflection
waveguides with periodic cladding, the band order $\nu$ defines a
number of zero crossings in every pair of the constitutive layers of
the cladding. For an aperiodic configuration of cladding this
condition is not satisfied, because some of the constitutive layers
appear to be repeated in a row producing a layer with a doubled
thickness. As a result, there are $\nu$, $\nu+2$ or $\nu+4$ zero
crossings in every pair of the structure layers. At the same time,
by analogy with a conventional slab waveguide \cite{snyder_1983},
the mode index $m$ specifies an integer number of the field's zeros
in the guiding layer. On the other hand an integer number of the
field peaks in the guiding layer can be defined as $m+1$. In this
notation, the even integer numbers correspond to the symmetric
modes, while the odd ones correspond to the anti-symmetric modes.

The lowest mode order of a conventional waveguide is usually $m =
0$. However as it is shown in \cite{li_josa2007}, when the thickness
of the guiding layer satisfies the condition
\begin{equation}
\label{eq:NegativeOrder} d_g \le d_2,
\end{equation}
where $d_2$ is the thickness of the second layer of the cladding, a
few modes with negative integer number $m=-1$ may emerge. In the
considered case $d_2 = 2d_\Upsilon = 4d_g$ and $d_2 = d_\Upsilon =
2d_g$ for aperiodic and periodic cladding, respectively, so
condition (\ref{eq:NegativeOrder}) is satisfied. The dispersion
curves of the TE$_{-1}$ mode are distinguished in
Fig.~\ref{fig:fig_2} by grey color. Both the number of field peaks
and zero crossings within the guiding layer become to be zero for
these modes. In fact, these modes can be interpreted as surface
modes whose energy is confined around the interface of two coupled
periodic/aperiodic structures \cite{VillaOE_2004}.

\begin{figure}[h]
\begin{center}
\begin{tabular}{c}
\includegraphics[height=5.0cm]{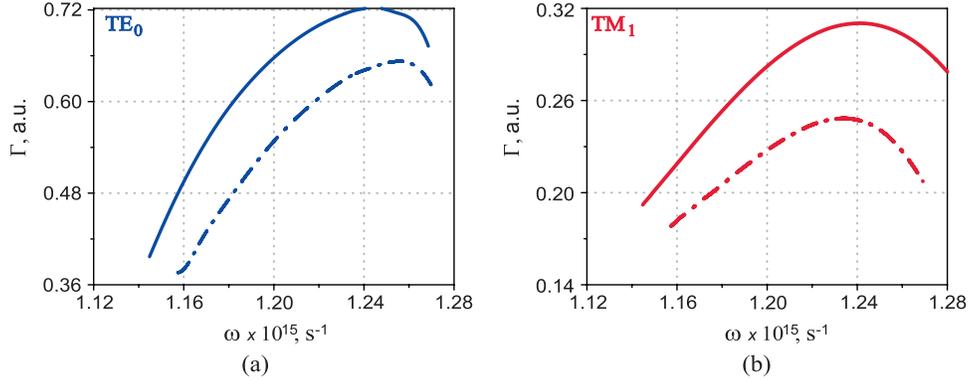}
\end{tabular}
\end{center}
\caption[example] {\label{fig:fig_5} Variation of the confinement
factor $\Gamma$ versus frequency $\omega$ for several guided modes:
(a) TE$_0$ modes; (b)TM$_1$ modes. Here solid curves correspond to
periodic cladding, while dashed curves correspond to aperiodic
cladding. In all cases $N=10$.}
\end{figure}

The confinement factor of several low-order modes is shown in
Fig.~\ref{fig:fig_5}. From these curves it follows, that the
confinement factor of both TM$_m$ and TE$_m$ modes is lower for
aperiodic structure as compared to the periodic one, so energy
leakage from the aperiodic cladding is greater. This is due to the
fact that bandgaps are narrower in the aperiodic systems. Although
the confinement factor of the modes is not high enough, the light
energy is confined strongly within the first few pairs of the
constitutive layers of the cladding as it is depicted in
Figs.~\ref{fig:fig_3} and \ref{fig:fig_4}.

Moreover, the confinement factor reduces, but remains finite, when
the dispersion curve approaches the band edges where there is a mode
cutoff. As a result, the confinement factor attains some maximum
between these two cutoff points. At the same time the confinement
factor of each mode is higher in the bandgap with the highest band
order $\nu$. This is due to the fact that the effective refractive
index $n_{eff}$ (which lies between 0 and $n_g$) acquires its
maximal value in this case. Note, in Bragg reflection waveguides a
larger confinement factor always corresponds to a larger value of
$n_{eff}=\beta/k_0$, because the normalized propagation constant
$\tilde \beta=(n_{eff}^2-n_{\Upsilon}^2)/(n_{\Psi}^2 -
n_{\Upsilon}^2)$ increases as  $n_{eff}$ rises resulting in the mode
field concentrating inside the core.

We should note, that in conventional waveguides based on the total
internal reflection \cite{snyder_1983} there is only a \emph{single}
cutoff point for each guided mode and $\Gamma$ gradually approaches
to 1 when operating frequency increases. Unfortunately it is
accompanied by increasing in the number of guided modes that become
to be propagating at this frequency. As a result, in such waveguides
with discrete spectra of guided modes, a confinement factor of the
fundamental mode approaches to its maximal value in the region where
high order modes also exist. In contrast to a conventional
waveguide, as it was demonstrated in previous section, Bragg
reflection waveguides and especially Bragg reflection waveguides
with aperiodic cladding are characterized by more complicated
spectra of guided modes. So that, in a certain frequency range, they
can support a single-mode regime with high degree of confinement of
the guided mode not only for the fundamental mode but also for other
high-order ones.

\subsection{Chromatic Dispersion}
\label{sec:GVD} Chromatic dispersion, which is manifested by the
frequency dependence of the phase velocity and group velocity on
refractive indexes of waveguide constituents (material dispersion)
as well as its geometrical parameters (waveguide dispersion), is a
significant factor in design of the photonic devices. In the field
of optical waveguides, a dispersion parameter $D$ is commonly used
for accounting effects of the chromatic dispersion. The dispersion
parameter $D$ may be calculated as \cite{Agrawal2001}:
\begin{equation}
\label{eq:GVD_} D = -\frac{\omega^2}{2\pi c} \beta'_2,
\end{equation}
where $\beta'_2 = d^2\beta'/d\omega^2$ is the group velocity
dispersion which is responsible for pulse broadening.

\begin{figure}[h]
\begin{center}
\begin{tabular}{c}
\includegraphics[height=5.0cm]{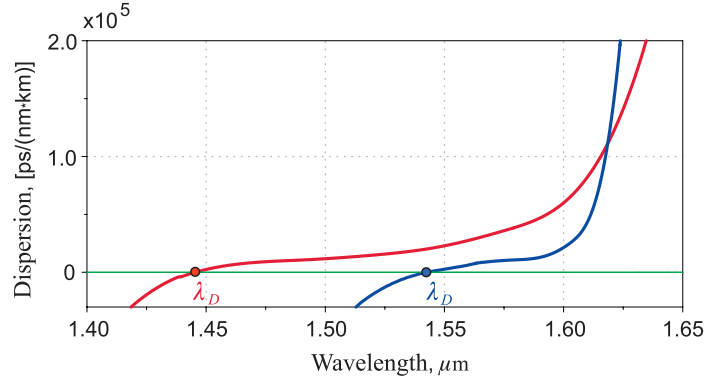}
\end{tabular}
\end{center}
\caption[example] {\label{fig:fig_6} Dispersion of the fundamental
TE$_0$ mode. Red and blue curves correspond to two particular
configurations of a Bragg reflection waveguide with periodic and
aperiodic cladding, respectively. Green line represents level
$D=0$~[ps/(nm $\cdot$ km)].}
\end{figure}

The dispersion curves of the fundamental TE$_0$ mode for both
periodic and aperiodic configurations of a Bragg reflection
waveguide are presented in Fig~\ref{fig:fig_6}. These configurations
are supposed to have the same geometrical parameters as considered
in Section~\ref{sec:dispersion}, but cladding's layers are made of
Si and SiO${_2}$. The choice of such a combination of the
constitutive media is provoked by the practical interest to the
wavelength range around $1.55~\mu$m, which is a telecommunication
standard.

We can see in Fig.~\ref{fig:fig_6} that the dispersion curves for
both configurations are quite similar to each other. Thus, the
dispersion parameter $D$ is positive and riches rather large values
almost in the whole bandwidth over $\lambda_D $ (here $\lambda_D$ is
the zero-dispersion wavelength) for both configurations. Moreover,
the dispersion parameter is flattened over a relatively large
wavelength range. When the dispersion curves of the corresponding
mode approaches the band edges (see, Fig.~\ref{fig:fig_2}) at large
frequencies, i.e. the shortest wavelengths in Fig.~\ref{fig:fig_6},
there are small regions with negative dispersion $D < 0$. Besides,
the dispersion parameter spikes in the long wavelength limit, where
$n_{eff}\to 0$ and $\beta' \to 0$.

At the same time, there are some distinctions in these dispersion
curves for periodic and aperiodic configurations, which are affected
by the waveguide dispersion contributions to $D$. Thus there is a
shift of $\lambda_D$  toward longer wavelengths at $\sim100$~nm for
aperiodic configuration. Here, $\lambda_D=1.44~\mu$m and
$\lambda_D=1.54~\mu$m are for Bragg reflection waveguides with
periodic and aperiodic cladding, respectively. Also we can conclude
that, for the waveguide with aperiodic cladding, flattened part of
dispersion curve is significantly shorter, compared to the case of
periodic cladding.

%%%%%%%%%%%%%%%%%%%%%%%%%%%%%%%%%%%%%%%%%%%%%%%%%%%%%%%%%%%%
\section{Conclusion}

In conclusion, we have proposed a novel type of a planar Bragg
reflection waveguide  which consists of an air guiding layer
sandwiched between two aperiodic mirrors which are arranged
according to the Kolakoski $K(1, 2)$ substitutional rule. On the
basis of the transfer matrix formalism, the bandgap conditions,
dispersion characteristics and mode profiles of the guided modes of
such Bragg reflection waveguide are studied. Peculiarities of the
dispersion properties of such Bragg reflection waveguide are
investigated.

We argue that the design of cladding in the form of an aperiodic
structure gives rise to changes in the waveguide dispersion,
resulting in the shift of zero-dispersion point toward the longer
wavelength. Besides, the flattened-dispersion region is more shorter
for the waveguide with aperiodic cladding than for periodic one

Since most of the characteristics of the planar Bragg waveguides are
similar to those of the cylindrical Bragg fibers, we argue that
obtained results are also applicable for prediction of optical
features of the latter ones.

%%%%%%%%%%%%%%%%%%%%%%%%%%%%%%%%%%%%%%%%%%%%%%%%%%%%%%%%%%%%%
\section*{Acknowledgments}

This work was partially supported (IAS and OVS) by University of
Guanajuato (projects DAIP-633/2015 and DAIP-609/2015).

\end{document}